
\def\unredoffs{\hoffset-.14truein\voffset-.2truein}
\def\redoffs{\voffset=-.55truein\hoffset=-.1truein} \def\speclscape{}
%
%
\newbox\leftpage \newdimen\fullhsize \newdimen\hstitle \newdimen\hsbody
\tolerance=1000\hfuzz=2pt
\catcode`\@=11 
\def\bigans{b }
\message{ big or little (b/l)? }\read-1 to\answ
\ifx\answ\bigans\message{(This will come out unreduced.}
\magnification=1200\unredoffs\baselineskip=16pt plus 2pt minus 1pt
\hsbody=\hsize \hstitle=\hsize 
\else\message{(This will be reduced.} \let\l@r=L
\magnification=1000\baselineskip=16pt plus 2pt minus 1pt \vsize=7truein
\redoffs\hstitle=8truein\hsbody=4.75truein\fullhsize=10truein\hsize=\hsbody
\output={\ifnum\pageno=0 
  \shipout\vbox{\speclscape{\hsize\fullhsize\makeheadline}
    \hbox to \fullhsize{\hfill\pagebody\hfill}}\advancepageno
  \else
  \almostshipout{\leftline{\vbox{\pagebody\makefootline}}}\advancepageno
  \fi}
\def\almostshipout#1{\if L\l@r \count1=1 \message{[\the\count0.\the\count1]}
      \global\setbox\leftpage=#1 \global\let\l@r=R
 \else \count1=2
  \shipout\vbox{\speclscape{\hsize\fullhsize\makeheadline}
      \hbox to\fullhsize{\box\leftpage\hfil#1}}  \global\let\l@r=L\fi}
\fi
%
\newcount\yearltd\yearltd=\year\advance\yearltd by -1900
%
\def\Title#1#2{\nopagenumbers\abstractfont\hsize=\hstitle
\centerline{\titlefont #2}\abstractfont\vskip .5in\pageno=0} %
%

\def\draftmode{\message{ DRAFTMODE }\def\draftdate{{\rm preliminary draft:
\number\month/\number\day/\number\yearltd\ \ \hourmin}}%
\headline={\hfil\draftdate}\writelabels\baselineskip=20pt plus 2pt minus 2pt
 {\count255=\time\divide\count255 by 60 \xdef\hourmin{\number\count255}
  \multiply\count255 by-60\advance\count255 by\time
  \xdef\hourmin{\hourmin:\ifnum\count255<10 0\fi\the\count255}}}
\def\nolabels{\def\wrlabeL##1{}\def\eqlabeL##1{}\def\reflabeL##1{}}
\def\writelabels{\def\wrlabeL##1{\leavevmode\vadjust{\rlap{\smash%
{\line{{\escapechar=` \hfill\rlap{\sevenrm\hskip.03in\string##1}}}}}}}%
\def\eqlabeL##1{{\escapechar-1\rlap{\sevenrm\hskip.05in\string##1}}}%
\def\reflabeL##1{\noexpand\llap{\noexpand\sevenrm\string\string\string##1}}}
\nolabels
%
\global\newcount\secno \global\secno=0
\global\newcount\meqno \global\meqno=1
\def\newsec#1{\global\advance\secno by1\message{(\the\secno. #1)}
\global\subsecno=0\eqnres@t\noindent{\bf\the\secno. #1}
\writetoca{{\secsym} {#1}}\par\nobreak\medskip\nobreak}
\def\eqnres@t{\xdef\secsym{\the\secno.}\global\meqno=1\bigbreak\bigskip}
\def\sequentialequations{\def\eqnres@t{\bigbreak}}\xdef\secsym{}
\global\newcount\subsecno \global\subsecno=0
\def\subsec#1{\global\advance\subsecno by1\message{(\secsym\the\subsecno. #1)}
\ifnum\lastpenalty>9000\else\bigbreak\fi
\noindent{\it\secsym\the\subsecno. #1}\writetoca{\string\quad
{\secsym\the\subsecno.} {#1}}\par\nobreak\medskip\nobreak}
\def\appendix#1#2{\global\meqno=1\global\subsecno=0\xdef\secsym{\hbox{#1.}}
\bigbreak\bigskip\noindent{\bf Appendix #1. #2}\message{(#1. #2)}
\writetoca{Appendix {#1.} {#2}}\par\nobreak\medskip\nobreak}
%
%
\def\eqnn#1{\xdef #1{(\secsym\the\meqno)}\writedef{#1\leftbracket#1}%
\global\advance\meqno by1\wrlabeL#1}
\def\eqna#1{\xdef #1##1{\hbox{$(\secsym\the\meqno##1)$}}
\writedef{#1\numbersign1\leftbracket#1{\numbersign1}}%
\global\advance\meqno by1\wrlabeL{#1$\{\}$}}
\def\eqn#1#2{\xdef #1{(\secsym\the\meqno)}\writedef{#1\leftbracket#1}%
\global\advance\meqno by1$$#2\eqno#1\eqlabeL#1$$}
%
\newskip\footskip\footskip14pt plus 1pt minus 1pt 
\def\footnotefont{\ninepoint}\def\f@t#1{\footnotefont #1\@foot}
\def\f@@t{\baselineskip\footskip\bgroup\footnotefont\aftergroup\@foot\let\next}
\setbox\strutbox=\hbox{\vrule height9.5pt depth4.5pt width0pt}
\global\newcount\ftno \global\ftno=0
\def\foot{\global\advance\ftno by1\footnote{$^{\#\the\ftno}$}}
%
\newwrite\ftfile
\def\footend{\def\foot{\global\advance\ftno by1\chardef\wfile=\ftfile
$^{\the\ftno}$\ifnum\ftno=1\immediate\openout\ftfile=foots.tmp\fi%
\immediate\write\ftfile{\noexpand\smallskip%
\noexpand\item{f\the\ftno:\ }\pctsign}\findarg}%
\def\footatend{\vfill\eject\immediate\closeout\ftfile{\parindent=20pt
\centerline{\bf Footnotes}\nobreak\bigskip\input foots.tmp }}}
\def\footatend{}
%
%
\global\newcount\refno \global\refno=1
\newwrite\rfile
\def\ref{[\the\refno]\nref}
\def\nref#1{\xdef#1{[\the\refno]}\writedef{#1\leftbracket#1}%
\ifnum\refno=1\immediate\openout\rfile=refs.tmp\fi
\global\advance\refno by1\chardef\wfile=\rfile\immediate
\write\rfile{\noexpand\item{#1\ }\reflabeL{#1\hskip.31in}\pctsign}\findarg}
\def\findarg#1#{\begingroup\obeylines\newlinechar=`\^^M\pass@rg}
{\obeylines\gdef\pass@rg#1{\writ@line\relax #1^^M\hbox{}^^M}%
\gdef\writ@line#1^^M{\expandafter\toks0\expandafter{\striprel@x #1}%
\edef\next{\the\toks0}\ifx\next\em@rk\let\next=\endgroup\else\ifx\next\empty%
\else\immediate\write\wfile{\the\toks0}\fi\let\next=\writ@line\fi\next\relax}}
\def\striprel@x#1{} \def\em@rk{\hbox{}}
\def\lref{\begingroup\obeylines\lr@f}
\def\lr@f#1#2{\gdef#1{\ref#1{#2}}\endgroup\unskip}

\def\addref#1{\immediate\write\rfile{\noexpand\item{}#1}} 
\def\startrefs#1{\immediate\openout\rfile=refs.tmp\refno=#1}
\def\xref{\expandafter\xr@f}\def\xr@f[#1]{#1}
\def\refs#1{\count255=1[\r@fs #1{\hbox{}}]}
\def\r@fs#1{\ifx\und@fined#1\message{reflabel \string#1 is undefined.}%
\nref#1{need to supply reference \string#1.}\fi%
\vphantom{\hphantom{#1}}\edef\next{#1}\ifx\next\em@rk\def\next{}%
\else\ifx\next#1\ifodd\count255\relax\xref#1\count255=0\fi%
\else#1\count255=1\fi\let\next=\r@fs\fi\next}
%

%
\newwrite\ffile\global\newcount\figno \global\figno=1
\def\fig{fig.~\the\figno\nfig}
\def\nfig#1{\xdef#1{fig.~\the\figno}%
\writedef{#1\leftbracket fig.\noexpand~\the\figno}%
\ifnum\figno=1\immediate\openout\ffile=figs.tmp\fi\chardef\wfile=\ffile%
\immediate\write\ffile{\noexpand\medskip\noexpand\item{Fig.\ \the\figno. }
\reflabeL{#1\hskip.55in}\pctsign}\global\advance\figno by1\findarg}
\def\xfig{\expandafter\xf@g}\def\xf@g fig.\penalty\@M\ {}
\def\figs#1{figs.~\f@gs #1{\hbox{}}}
\def\f@gs#1{\edef\next{#1}\ifx\next\em@rk\def\next{}\else
\ifx\next#1\xfig #1\else#1\fi\let\next=\f@gs\fi\next}
\newwrite\lfile
{\escapechar-1\xdef\pctsign{\string\%}\xdef\leftbracket{\string\{}
\xdef\rightbracket{\string\}}\xdef\numbersign{\string\#}}

\def\writestop{\def\writestoppt{\immediate\write\lfile{\string\pageno%
\the\pageno\string\startrefs\leftbracket\the\refno\rightbracket%
\string\def\string\secsym\leftbracket\secsym\rightbracket%
\string\secno\the\secno\string\meqno\the\meqno}\immediate\closeout\lfile}}
\def\writestoppt{}\def\writedef#1{}
\def\seclab#1{\xdef #1{\the\secno}\writedef{#1\leftbracket#1}\wrlabeL{#1=#1}}
\def\subseclab#1{\xdef #1{\secsym\the\subsecno}%
\writedef{#1\leftbracket#1}\wrlabeL{#1=#1}}
\newwrite\tfile \def\writetoca#1{}
\def\leaderfill{\leaders\hbox to 1em{\hss.\hss}\hfill}
\def\writetoc{\immediate\openout\tfile=toc.tmp
   \def\writetoca##1{{\edef\next{\write\tfile{\noindent ##1
   \string\leaderfill {\noexpand\number\pageno} \par}}\next}}}

%
\catcode`\@=12 
%
\edef\tfontsize{\ifx\answ\bigans scaled\magstep3\else scaled\magstep4\fi}
\font\titlerm=cmr10 \tfontsize \font\titlerms=cmr7 \tfontsize
\font\titlermss=cmr5 \tfontsize \font\titlei=cmmi10 \tfontsize
\font\titleis=cmmi7 \tfontsize \font\titleiss=cmmi5 \tfontsize
\font\titlesy=cmsy10 \tfontsize \font\titlesys=cmsy7 \tfontsize
\font\titlesyss=cmsy5 \tfontsize \font\titleit=cmti10 \tfontsize
\font\titlebfsl=cmbxsl10 \tfontsize
\skewchar\titlei='177 \skewchar\titleis='177 \skewchar\titleiss='177
\skewchar\titlesy='60 \skewchar\titlesys='60 \skewchar\titlesyss='60
\def\titlefont{\def\rm{\fam0\titlebfsl}
\textfont0=\titlerm \scriptfont0=\titlerms \scriptscriptfont0=\titlermss
\textfont1=\titlei \scriptfont1=\titleis \scriptscriptfont1=\titleiss
\textfont2=\titlesy \scriptfont2=\titlesys \scriptscriptfont2=\titlesyss
\textfont\itfam=\titleit \def\it{\fam\itfam\titleit}\rm}
 \ifx\answ\bigans\else scaled\magstep1\fi
\ifx\answ\bigans\def\abstractfont{\tenpoint}\else
\font\abssl=cmsl10 scaled \magstep1
\font\absrm=cmr10 scaled\magstep1 \font\absrms=cmr7 scaled\magstep1
\font\absrmss=cmr5 scaled\magstep1 \font\absi=cmmi10 scaled\magstep1
\font\absis=cmmi7 scaled\magstep1 \font\absiss=cmmi5 scaled\magstep1
\font\abssy=cmsy10 scaled\magstep1 \font\abssys=cmsy7 scaled\magstep1
\font\abssyss=cmsy5 scaled\magstep1 \font\absbf=cmbx10 scaled\magstep1
\skewchar\absi='177 \skewchar\absis='177 \skewchar\absiss='177
\skewchar\abssy='60 \skewchar\abssys='60 \skewchar\abssyss='60
\def\abstractfont{\def\rm{\fam0\absrm}
\textfont0=\absrm \scriptfont0=\absrms \scriptscriptfont0=\absrmss
\textfont1=\absi \scriptfont1=\absis \scriptscriptfont1=\absiss
\textfont2=\abssy \scriptfont2=\abssys \scriptscriptfont2=\abssyss
\textfont\itfam=\bigit \def\it{\fam\itfam\bigit}\def\footnotefont{\tenpoint}%
\textfont\slfam=\abssl \def\sl{\fam\slfam\abssl}%
\textfont\bffam=\absbf \def\bf{\fam\bffam\absbf}\rm}\fi
\def\tenpoint{\def\rm{\fam0\tenrm}
\textfont0=\tenrm \scriptfont0=\sevenrm \scriptscriptfont0=\fiverm
\textfont1=\teni  \scriptfont1=\seveni  \scriptscriptfont1=\fivei
\textfont2=\tensy \scriptfont2=\sevensy \scriptscriptfont2=\fivesy
\textfont\itfam=\tenit \def\it{\fam\itfam\tenit}\def\footnotefont{\ninepoint}%
\textfont\bffam=\tenbf \def\bf{\fam\bffam\tenbf}\def\sl{\fam\slfam\tensl}\rm}
\font\ninerm=cmr9 \font\sixrm=cmr6 \font\ninei=cmmi9 \font\sixi=cmmi6
\font\ninesy=cmsy9 \font\sixsy=cmsy6 \font\ninebf=cmbx9
\font\nineit=cmti9 \font\ninesl=cmsl9 \skewchar\ninei='177
\skewchar\sixi='177 \skewchar\ninesy='60 \skewchar\sixsy='60
\def\ninepoint{\def\rm{\fam0\ninerm}
\textfont0=\ninerm \scriptfont0=\sixrm \scriptscriptfont0=\fiverm
\textfont1=\ninei \scriptfont1=\sixi \scriptscriptfont1=\fivei
\textfont2=\ninesy \scriptfont2=\sixsy \scriptscriptfont2=\fivesy
\textfont\itfam=\ninei \def\it{\fam\itfam\nineit}\def\sl{\fam\slfam\ninesl}%
\textfont\bffam=\ninebf \def\bf{\fam\bffam\ninebf}\rm}
%
%

\hyphenation{anom-aly anom-alies coun-ter-term coun-ter-terms}
\def\inv{^{\raise.15ex\hbox{${\scriptscriptstyle -}$}\kern-.05em 1}}

\def\Dsl{\,\raise.15ex\hbox{/}\mkern-13.5mu D} 
\def\dsl{\raise.15ex\hbox{/}\kern-.57em\partial}

\font\bigit=cmti10 scaled \magstep1
\def\lspace{\ifx\answ\bigans{}\else\qquad\fi}
\def\lbspace{\ifx\answ\bigans{}\else\hskip-.2in\fi} 
\def\boxeqn#1{\vcenter{\vbox{\hrule\hbox{\vrule\kern3pt\vbox{\kern3pt
        \hbox{${\displaystyle #1}$}\kern3pt}\kern3pt\vrule}\hrule}}}
\def\mbox#1#2{\vcenter{\hrule \hbox{\vrule height#2in
                \kern#1in \vrule} \hrule}}  
%

\def\darr#1{\raise1.5ex\hbox{$\leftrightarrow$}\mkern-16.5mu #1}

\def\roughly#1{\raise.3ex\hbox{$#1$\kern-.75em\lower1ex\hbox{$\sim$}}}
\def\daitai{\lower.3ex\hbox{$-$\kern-.75em\raise.6ex\hbox{$\sim$}}}

%
\def\AP#1{Ann.\ Phys. {\bf{#1}}}

\def\NP#1{Nucl.\ Phys. {\bf B{#1}}}
\def\PL#1{Phys.\ Lett. {\bf B{#1}}}

\def\PRD#1{Phys.\ Rev. {\bf D{#1}}}

\def\PRL#1{Phys.\ Rev.\ Lett. {\bf {#1}}}
\def\PTP#1{Prog.\ Theor.\ Phys. {\bf {#1}}}

\message{ This is KUMACS.TEX ver.1.1,}
\message{shamelessly stolen from harvmac.tex,}
\message{a bit modified and unnecessarily (re-)named}
\message{by K.Harada}
%
\input ptkumacs.tex


\def \Rn#1 {\uppercase\expandafter{\romannumeral#1}}
\def\gtsim{\mathrel{\hbox{\raise0.2ex
\hbox{$>$}\kern-0.75em\raise-0.9ex\hbox{$\sim$}}}}
\def\ltsim{\mathrel{\hbox{\raise0.2ex
\hbox{$<$}\kern-0.75em\raise-0.9ex\hbox{$\sim$}}}}
\let \q=`
\def \Romannumeral(#1) {\uppercase\expandafter{\romannumeral#1}}

\def \Romannumeral(#1) {\uppercase\expandafter{\romannumeral#1}}
\def \Rn(#1) {\uppercase\expandafter{\romannumeral#1}}
\def\Fig(#1){$${\overline{\underline{\rm Fig.{\ \ #1}}}}$$} 
\def\Tab(#1){$${\overline{\underline{
 \rm Table\ \  \uppercase\expandafter{\romannumeral#1}}}}$$} 
\def \ds{\displaystyle}
\def \a{\alpha}
\def \b{\beta}
\def \g{\gamma}
\def \c {\chi}

\def \ve{\varepsilon}
\def \l {\lambda}
\def \m{\mu}
\def \n{\nu}
\def \p{\psi}

\def \t{\tau}

\def \th{\theta}

\def \c {\chi}

\def \l {\lambda}
\def \p{\psi}


\def \Ft{\tilde F}

\def \fn1 {N_{f_1}}
\def \fn2 {N_{f_2}}

\def \Romannumeral(#1) {\uppercase\expandafter{\romannumeral#1}}
\def \Rn(#1) {\uppercase\expandafter{\romannumeral#1}}

\vskip 4cm
\Title{KYUSHU-HET-20, SAGA-HE-72}
{\vbox{\centerline{}
\vskip0.2em\centerline{}
\vskip0.2em\centerline{}
\vskip0.2em\centerline{}
\vskip0.2em\centerline{}
\vskip0.2em\centerline{ Real Space Renormalization Group Analysis }
\vskip0.2em\centerline{of U(1)-Gauge Theory}
\vskip0.2em\centerline{ with  $\th $ term in 2 Dimensions}}}
\centerline{ Ahmed Sayed HASSAN, Masahiro IMACHI}
\smallskip
\centerline{Department of Physics,Kyushu University,Fukuoka, 812 JAPAN}
\smallskip
\centerline{and}
\smallskip
\centerline{Hiroshi YONEYAMA}
\centerline{ Department of Physics,Saga University, Saga, 840 JAPAN}
\vskip1.0cm
\centerline{{\bf Abstracts}}
 U(1) lattice gauge theory with $\th$-term is investigated by real space
 renormalization group approach. Flows of renormalized coupling constants are
analyzed. For each $\th$, renormalization flows converge to a single
trajectory irrespective of  bare coupling constants of real action.
For $\th \not= \pi$ the system is in the confinement phase
controlled by an infrared fixed point. We found a phase transition
at $\th = \pi$.
 Imaginary part of the action given by $\th$-term stays fixed under
renormalization group transformations leading to deconfinement.
\par
\newsec{Introduction} 
Much progress has been made in understanding gauge theories on lattice by
various approaches. The systems are, however, limited mainly to those without
$\th$-terms. It is due to the difficulty arising from the $\th$-term
which leads to imaginary action in the Euclidean space preventing it
to be analyzed by standard numerical simulations.
The nature of gauge theories with $\th$-terms lacks in understanding and
waiting for further extensive  studies.
\par
According to Polyakov \ref \Pol{A. M. Polyakov,\NP{120}(1977), 429.},
compact U(1) gauge theory in 3-dimensions is in
the confining phase due to monopole excitations. If the system contains
$\th$-term, does the system still stay  in the confining phase or not?\par
 This question is still unresolved. There is a conjecture that under the
existence of $\th$-term( Chern-Simons term), the system will be in
the deconfining phase due to the effect of $\th$-term, which  washes
out monopole excitations\ref \FS{ E. Fradkin and F. A. Shaposnik,
\PRL{66}(1991), 276.}.
Whether this conjecture is really realized in nature or not is waiting for
investigations.
\par
As a first step to approach to these problems,  we attempt, in this paper,
to study compact U(1) gauge system with $ \th$-term
 in 2-dimensions by real space renormalization group(RG) analysis
\ref\Mig{A. A. Migdal, Sov. Phys. JETP    {\bf 42}(1976) 413; 743.}
\ref \Kad  {L. P. Kadanoff,\AP { 100}(1976), 359.}
\ref \BGZa {K. M. Bitar, S. Gottlieb and C. K. Zachos,\PRD { 26} (1982), 2853.}
\ref\BGZb{K. M. Bitar, S. Gottlieb and C. K. Zachos, \PL { 121} (1983), 163.}
\ref\IKY {M. Imachi, S. Kawabe and H. Yoneyama,\PTP{ 69} (1983), 221; 1005.}.
 The study of 3-dimensions will be studied in the near future. \par
We study following points.
\par
\item{(i)} RG flows leading to universal renormalized trajectories for each
 $\th$.\par
\item{(ii)} The change in $\th$-term under RG transformations.
 Special value $\th=\pi$ gives a fixed point of RG transformation.\par
\newsec{RG transformation and RG flow} 
\noindent{\bf 2-1 RG flow }\par
\itemitem{(1)} We show renormalized trajectories in coupling constant space
for the real action (usual action ) and the imaginary action ( induced by
$\th$-term).
\par
\itemitem{(2)} The system is shown to lie in the confinement phase for
$\th ({\rm bare})\not= \pi$. \par
\itemitem{(3)} It lies in the deconfinement phase for $\th({\rm bare})=\pi$.
The imaginary part of action with the form $i {\th \over 2\pi}F_{10}$
stays at fixed point at $\th=\pi$. The real part of the action moves to
a fixed point starting from various Wilson type bare actions. \par
The partition function in 2-dimensions is expressed as
$$ \eqalign{
            Z(L)=&{1\over 2\pi}\int dx F(L,x) \cr
           F(L, x)=&\exp \{-s(L, x)+i v(L, x) \} }\eqno(1) $$
where $L$ and $x$ denotes the scale and the field strength $ F_{\m \n } $
respectively. Function $F$ in eq.(1) will be called as F-function hereafter.
Functions $s(L,x)$ and $v(L,x)$ are the real (ordinary Euclidean) action
and the imaginary
 action induced by the $\th$-term.
The bare $\th$-term, i.e., the imaginary action at $L=a=$ lattice constant
gives $v(a,x)=-{\th \over {2\pi}} F_{0 1}$, which will be simply written as
$-{\th \over {2\pi}} x $. Under  RG transformations, $v(L, x)$ at larger
length scales will have higher
power terms in  $x$,
and its functional form is determined by RG transformations.
\par
The action changes under RG transformations. The action is expanded as a
linear combination of irreducible(IR) characters of the
gauge group. The real  and the imaginary action $s$, and $i v$ are
$$ s=\sum_{q=-\infty}^{\infty} \b_q \chi_q(x) $$
$$  v=  \sum_{q=-\infty}^{\infty}\g_q \chi_q(x) \eqno(2)$$
where $\beta_q$ and $\gamma_q$ are coupling constants. The function
$ \chi_q (x)=\exp (i q x)$ is the character of $q$-IR representation
($q=$ integer) and $x$ denotes the field strength.
For example the Wilson action in pure gauge without $\th$-term is given by
\par
$$\eqalign{
  \b_1 =&{1\over g^2},{\rm \hskip 3cm} \b_q=0 {\hskip 1cm} (q \not= 1), \cr
 \g_q=&0  {\rm {\hskip 2cm} for}{\hskip 1cm} q={\rm all}.} \eqno(3)$$
When the system contains $\th$-term, typical $\th$-term is given by
\par
$$ v={\th \over {2\pi}} x
           =i{\th \over {2\pi}}\sum_{q=-\infty}^{\infty}{(-1)^q\over q}
            \chi_q(x) \eqno(4)$$
where $x=F_{0 1}$ and thus
$$\g_q=i {\th \over{2\pi}}{(-1)^q \over q} \hskip 1cm{\rm for \ all}\ \ \
                         q( q\not=0 ).
                                                   \eqno(5)$$
For the real action $s$, the gaussian form(=Maxwell term
=$-{2\over g^2}x^2=-{2\over g^2}F_{\m \n}^2 $) is expanded with coefficients
\par
$$ \b_q=-{{2(-1)^q} \over q^2} {1 \over g^2} \hskip 1cm({\rm for \ all}
                                 \ \ \ q)  \eqno(6)$$
\vskip 1cm
\noindent {\bf RG transformation }\par
Starting from some bare action, the transformed
action is given through the relation between F-functions $F(L, x)$ and
$F(\lambda L, x)$,
\par
$$ F(L, x)=\exp\{-s(L, x)+i \ v(L,x)\}, $$
$$F(\l L, x)=\exp\{-s(\l L, x)+i \ v(\l L,x)\},      \eqno(7)$$
where $s$ and $v$ are defined by the coupling constants at scale
$L$ in $F(L,x) $ and at $\l L$ in $F(\l L,x)$,  respectively. \par
%
Recursion equation is given by
$$  F(\l L,x)=\sum_{q} \c_q(x) \Ft_q^{\l^2} (L) \eqno(8)$$
where $\Ft_q (L)$ is defined by the IR character expansion of
 $F(L,x) $ at scale $L$ as  $F(L,x)=\sum_{q} \c_q(x) \Ft_q (L) $.\par
Using this recursion equation (8), we can calculate the coupling constants
at larger scales. The analytic properties derived from the recursion equation
will be discussed in section 3( ``Convolution") of this paper.\par
Free energy is given by
$${1 \over L^2}(-\ln Z(L))
= {1\over L^2}(-\ln {\tilde F}_0 (L)) ={1\over {\l^2 L^2}}(-\ln
{\tilde F_0 }^{\l^2} (L))$$
at any scale $L$,
where $Z(L)=\int {dx \over 2\pi} F(L, x)={\tilde F}_0 (L)$.  \par
%
\noindent{\bf Renormalized Trajectories }\par
Flows in coupling space under RG transformations are calculated
 for  bare actions with various bare $\th$-terms $(\theta \not= 0, \pi)$.
Under  RG transformations,  a set of coupling constants moves
in the infinite dimensional coupling constant space. Starting from
various bare  real actions for a fixed $\th$, we clearly see that they
converge to a single trajectory
 for each  imaginary ($ \th $) action. The universal trajectory is called
renormalized trajectory. Since we have extended the coupling constant space to
real ($\b _q$'s) and imaginary ($\g_q$'s),  we have doubly
infinite dimensional
 coupling constant space. \par
\item{(i)} For each $\th$, various actions with different bare coupling
constants $\b^b$'s converge
to a unique trajectory( Fig. 1 and Fig. 2) after 2 or 3 steps of RG
 transformations. Unique trajectories are defined for each $\th$. Unique
trajectories with different bare  $\th^b$'s differ from each others. But they
converge to a
$\th$ independent single trajectory in
intermediate or strong
 coupling regions( $|\g| \leq |\b|\ltsim 1/2$ ).
The behavior of real coupling constants $\b_q$'s are controlled by the
imaginary action( $\th$-term ). In weak coupling regions, real couplings run
along Gaussian trajectory( $\propto x^2=F_{\m \n}^2$ ) as shown in Fig. 1. \par
\Fig(1)
\Fig(2)
To see more closely the universality,
we show $\{\beta _q \}$ and $\{{\rm Im} \g_q \}$ in the strong coupling region
 in Fig.3 and Fig. 4 respectively.
Various bare actions starting from various $\b^b$'s lead to quite
clearly to a universal trajectory.
\Fig (3)
\Fig (4)  
The exceptional case is ${\th =\pi}$. The imaginary action does not
 change and the real action is affected by the stationary imaginary action.
 The flow  converges to a point  $(\b_1, \b_2)=( {1 \over 2}, -{1\over 4})$
 as shown in Fig. 5.\par
\Fig(5) 
%
In these figures, infinite dimensional coupling constant space is projected
onto
  $(\beta_1, \beta_2)$, and
$(\gamma_1, \gamma_2)$ plane.\par
For $\th^b$ very close to $\pi$, e.g., $\th \approx0.999\pi $, $\b_q $
started from weak coupling regions moves along Gaussian line in $\b_q$ space
and reaches  the intermediate regions where $\b_q$'s  stay for many steps
near the fixed point $\b_q =(-1)^{q-1}/2 |q|$ (for all $q$). It then moves
 along the trajectory controlled by $\th$-term to the infrared fixed point(
$\{\b_q\}=0)$
(see Fig. 6(a)).\par
\item{(ii)} Renormalized trajectory projected onto $(\g_1, \g_2)$ space shows
 that trajectory is almost independent of $\b_q$'s and moves on a unique
trajectory to $\g_q=0$ if we start from $\th^b \not=0, \pi$. The case $\th^b
 =\pi$
is exceptional. The $\th$-term stays at the same value( $\g_q= i (-1)^q /2q
( q\not= 0), \g_0=0$).\par
\item{(iii)} Renormalized trajectories projected onto $(\b_1, {\rm Im}\g_1)$
plane
provide very important information about the phase structures( Fig. 6(b)).
\Fig({6(a)(b)})
At $\b_1$ large, $\b_q$'s move toward strong coupling region along  each
trajectory defined by $\th^b$. In intermediate or strong coupling regions,
trajectories started with various $\th^b$'s converge to a single trajectory
controlled by $\th$ term.\par
 Based on this fact, we can calculate $t_G$-$\th^b$
relation by setting a ``gate" $\b_G $, where $t_G$ is the number of RG
transformations necessary to reach the gate from each bare coupling
constant.   The relation
$t_G$-$\th^b$ shows the phase transition at $\th=\pi$. This point will be
discussed in the following subsections.\par
\vskip 1cm
%
\noindent{\bf 2-2 Gell-Mann Low function}\par
Flow diagram of  the renormalization group trajectories of coupling constants
is quite useful to study the phase structure of the system. We observed
that  every trajectory starting from various bare  actions
converges to a unique trajectory. We choose a point, ``gate", $\b_G$ on
the universal renormalized trajectory(RT) and calculate the amount
of change of scale between the length
scale at the gate
$\b_G$ and that at the bare coupling $\b^b
\equiv \b(a)$. This calculation  gives us the relation between $\b(a) $ and
$a(=$ cutoff).
In Migdal-Kadanoff RG
 approach
( Note that the M-K RG is exact in two dimensions)
,
 the scale change is given by $\l$
( for which we
 take $\l ={ \sqrt 2}$ in this paper) and the ``physical" length $\xi _G$
at the gate is related to lattice cut-off $a$ as
$$ \xi_G =\l^{t_G} a                \eqno(9)$$
where $t_G$ is the number of RG transformations necessary to reach
$\b_G $ from $ \b^b=\b(a) $.\par
Since $\xi_G$ is a physical length scale, it should be independent of $a$.
 The quantity $t_G $ is calculated as  a function of bare coupling $\b^b=\b(a)$
 and the relation (9) gives us the functional relation between $\b^b$ and $a$.
 This relation is equivalent  to the  beta function of the renormalization
group. Gell-Mann Low function
$ \p(\b^b)$ (=beta function) is defined as
$$ \p(\b^b)=1/(d t_G/d \b^b)|_{\b_G}. \eqno(10)$$
In this paper we are interested in the phase structure of the $\th$-action. So
we calculate $t_G$-$\th^b$ relation. We calculated $t_G$ for various bare
$\th$-parameters( $\th^b$'s) for common $\b^b$. The result is shown in
Fig.(7) as  a function of $t_G(\th^b)$, where $\b_G $ is fixed at $10^{-3}$.
\Fig(7)
 We found that
there  is a phase transition at $\th^b =\pi \equiv  \th_c$ ( $c$ means
critical)
 since $t_G \rightarrow  \infty$ and $d t_G/d \b^b \rightarrow \infty$
at  $\th^b \sim \pi$ ( actually, the value of $t_G$ near $\th_b=\pi$ are
for example
 $t_G(\th^b=0.999 \pi)=t_G( \th^b=1.001  \pi)=12.12   )$.
%
%
%
\newsec{Change in $\th$-term under RG transformation and Fixed points}
%
We investigate the convolution of $\th$-action.
Now we consider the convolution of two plaquettes as shown in Fig. 8.
\Fig(8)
We will investigate how bare $\th$-term action transforms under RG.
%
The function
$ F(x)= e^{ i \alpha x} $ for single plaquette $(\a =\th / 2 \pi )$
can be expanded by IR characters:
$$F(x)= e^{i \a x}=\sum_q \chi_q(x) \Ft_q \eqno(11)$$
where
$$\Ft_q= { (-1)^q \over (\a-q)}   { \sin {\pi \a} \over \pi}.
\eqno(12)$$
 Consider two  neighboring plaquettes and integrate over the link variable
$x$ common to both plaquettes(Fig. 8). Is it given by the following equation?
$$ \int_{-\pi}^{\pi} {dx\over {2\pi}} F(x+y)  F(-x)=
\int {dx\over {2\pi}} e^{i \a (x+y)} e^{-i \a x}=e^{i \a y}.
 \eqno({\rm I}) $$
If so, $\theta$-term action does not change its functional form for any
 $\th (\a\equiv \th/2\pi)$ parameter. If (I) holds, it leads to a strange
result. From the l.h.s. of (I), we have
$$ \int_{-\pi}^{\pi} {dx\over {2\pi}} F(x+y)  F(-x)=
{\sum_{q q'}} \int_{-\pi}^{\pi} {dx \over {2\pi}}
    \chi_q(x+y)  {\chi^*_{q'}} (-x)\Ft_q   \Ft_{q'}$$
$$=\sum_q \chi_q(y) \Ft^2_q =\sum_q \chi_q(y)
{ 1  \over (\a-q)^2}   { \sin^2 {\pi \a} \over \pi^2},\eqno(13)$$
by IR character expansion.\par
This form is obviously different from r.h.s. of (11) since we have a
 coefficient $\Ft^2_q$ in (13) instead of $\Ft_q$ in r.h.s. of (11),
so the result
should be different from the last term in (I).
\par
The important property of $F$ is the periodicity, namely
%
$$F(x+y)=\cases {
                 \ds{ e^{ i \a   (x+y) }   {\hskip 2cm}    {\rm for}
                                     {\hskip 2cm} x=[-\pi, \pi -y] }\cr
           \noalign{\vskip 1cm}
                 \ds{ e^{ i \a   (x+y-2 \pi) } {\hskip 1.5cm} {\rm for}
                                {\hskip 2cm} x=[\pi -y, \pi ].}
                                                                }
                                                            \eqno(14)$$
where    $y \geq  0$ is assumed.
Then the correct convolution is
$$ \int F  F=\int_{-\pi}^{\pi-y} {dx \over 2\pi}e^{i \a (x+y)}
 e^{-i \a x}+\int_{\pi-y}^{\pi} {dx \over 2\pi} e^{i \a (x+y-2\pi)}
 e^{-i \a x}
  $$
$$ =e^{i \a y} \{ 1-{y\over 2\pi}(1-e^{- i  2\pi \a})\}, {\hskip 1cm}
{\rm for}  {\hskip 1cm} y>0
\eqno({\rm II})$$
similarly,
$$ \int F  F =e^{i \a y} \{ 1+{y\over 2\pi}(1-e^{+ i  2\pi \a})\}
 {\hskip 1cm} {\rm for}  {\hskip 1cm} y<0  \eqno({\rm II'})$$
which is evidently different from the last term of (I). From (II) and (II$')$
 we understand
that  $\th$-action will generally change due to complex
contribution in the bracket of r.h.s. of (II)'s, but when $\a =
  {1 \over 2}$ or  $\th = \pi $ the correction in bracket becomes purely
real and no modification appears in the imaginary($\th$ -term) action.
 It shows that the fixed point action appears only for $\th=\pi$.
\par
The remaining thing to verify is to show the equivalence between r.h.s.
of (II)
and
 r.h.s. of (13). \par
{}From
$${e^{i \a y}} =\sum_q \chi_q(y)  { (-1)^q  \over (\a-q)}
   { \sin {\pi \a} \over \pi}.\eqno(15)$$
we have
$${e^{i \a (y-\pi )}} =\sum_q \chi_q(y)  { 1 \over (\a-q)}
   { \sin {\pi \a} \over \pi}.\eqno(16)$$
 Differentiating (16) with respect to $\a$, we obtain
$$\sum_q \chi_q(y)  { 1  \over (\a-q)^2}= { \pi^2  \over
\sin^2 {\pi \a}} {e^{i \a y}} {1 \over {2\pi}} {(2\pi -y
+y {e^{-i  2\pi \a}})},  \eqno(17)$$
 which gives us the equality of r.h.s. of (II) and (13).
The infinite series (17) gives the r.h.s. of (II).
The partition function obtains a contribution in the imaginary
 action since
 $\theta$-parameter gives $e^{-i 2\pi\alpha} \not= {\rm real}$ for $\th\not=0,
\pi$.
The important observation about  eq.(II) is that
{\it  the special nontrivial value  $\th=\pi$
  keeps the form of $\th$-term action unchanged.
          Namely bare $\theta=\pi$ is the fixed point  of
  the RG transformation of the imaginary action.}\par
\noindent ($\theta=0$ is another fixed point of imaginary action. In this
 case total action is usual real one.)\par
Eq.(II) states that $\th$-parameters other than these special points give
change to imaginary action under  RG transformations, i.e., it will give RG
trajectory in $\{\g_q\}$ space.\par
The result that $\theta=\pi$ is a fixed point in RG is consistent with
(1) MC simulation ( Wiese)\ref \Wi {U. -J. Wiese, \NP{ 318}(1989) 153.}and
(2) classical arguments by Coleman\ref \Col  {S. Coleman,\AP{ 101}(1976) 239.}
 that $\theta=\pi$ is the special value which
leads to deconfinement in 2 dimensional QED( Schwinger Model), where
background electric field produced by the $\th$-term just cancels confining
force for $\th=\pi$.\par
\vskip 1cm
\newsec{Relation between real and imaginary($\th$) action in strong coupling
 regions} 
Except for the special case with fixed point ($\th=\pi$) the system
is controlled by the infrared fixed points at $\{ \b_q\}=\{\g_q\}=0$. So here
 we mainly
consider the region where $|\b_q|$'s and $|\g_q|$'s are small
 compared with unity. We denote $i \g_q$ as $\t_q$. Under RG transformation,
 $\b_1$ and $\t_1$ changes as
\par
$$\b_1(\l L)={\b_1^2(L)+\t_1^2(L) \over{1+2f(L)}},    \eqno(18) $$
$$\t_1(\l L)={2\b_1(L)\t_1(L) \over{1+2f(L)}},    \eqno(19) $$
where
$$f (L)= {\b_1}^2(L) +{\t_1}^2(L) + \b_1(L) \t_{-1}(L)+ \b_{-1}(L)
 \t_1(L)   \eqno(20)$$
with  ${\b_{-1}={\b_1}^*=\b_1}$=real, and
  ${\t_{-1}={-\t_1}^*=-\t_1}$=real. \par
We will list several properties of $\b_q$'s and $\t_q$'s.
\par
\item{(i)} If $\t_q(L)=0$ for all $q$, we have $\t_q(\l L)=0$ for all $q$.
 Namely, we will not encounter $\th$-term in pure gauge
system when bare action dose not contain $\th$-term.\par
\item{(ii)} When $|\t_q|$'s are much smaller than $|\b_q|$'s,
at some scale $L$ , we have yet much small $|\t_q|$'s at scale  $\l L$.
Since $ \t_q(\l L)  / \b_q(\l L) {\sim} 2 \ve $ when $\t_q(L) / \b_q(L) =
\ve(\ll 1) $,  $ \t_q(\l L) / \b_q(\l L)$
is a quantity of $O(\ve)$, but its magnitude is larger than
$\t_q(L)/ \b_q(L)$. Under many times of repetitions of scale transformation it
will grow larger.\par
\item{(iii)} If  ${\b_1(L) = \t_1(L)}$ (other $\b_q $'s and $ \t_q$'s
with $q \ge 2$ are
supposed to be very small ), we have $ \b_1(\l L) = \t_1(\l L)$.\par
\item{(iv)} Suppose that $\b_1(L)$ is different from $\t_1(L)$ and that $
\b_1 $
is close to $ \t_1$. Namely,
 $${\b_1(L) \over \t_1(L)}={1+k}, (k\ll 1) \eqno(21) $$
then
$$ \cases{
\ds{ \b_1(\l L) = 2 (1+k+ { k^2 \over 2 } ) {\t_1}^2(L)/ (1+f)    }\cr
   \noalign{\vskip .5cm}
\ds{ \t_1(\l L) = 2 (1+k) {\t_1}^2(L) / (1+f)   }
                                                    }\eqno(22) $$
So we have

$${\b_1(\l L) \over \t_1(\l L)}={ {1+k+{k^2 \over 2}} \over {1+k}}
                                                   =1+{k^2 \over 2}\eqno(23) $$
which is much closer to unity than $ \b_1(L) / \t_1(L)$.
\item{(v)} For ${\b_q^b} \equiv \b_q(a)=0 $ and  ${\t_q^b} \equiv \t_q(a)=$
small
 $\not=0 $ , we have $\b_q(L) \simeq \t_q(L) = $ small, and the flow rapidly
converges to the infrared fixed point $\{\b_q\} =\{ \t_q\} =0. $
\par
\item{(vi)} Exception to (v) is  the case with $\th = \pi $.
In this case $ \th $-term does not change under RG transformation, i.e.,
$ \t_q(L) =-\pi (-1)^q / q  $ ( for any $q(\not=0))$ at any scale $L$.
\par
Even for $\b_q^b=0$ , $\b_q(L)$ becomes non zero when $\th=\pi$.
 At the first step($t=1$)
in RG transformations, we have
$$\b_1( \l a) ={1 \over \pi} \int_{0}
 ^{ \pi} {\sin t \over t } dt = 0.58949... \eqno(24)$$
Since
$$ \b_1( \l a) = \int_{- \pi}
 ^{ \pi} {dx \over 2 \pi} { \chi _1^*(x) \ln (1-{|x| \over \pi})
                                    }\eqno(25)  $$
 from eq.(II), and
$$ F( \l a,x)= e^{ i {x \over 2}} (1-{|x| \over \pi}).\eqno(26)$$
After further RG transformation , we have $\b_1( {\l^t} a) \simeq
\t_1( {\l^t} a) \simeq {1 \over 2}$ ( fixed point value) due to property (iv).
Actually $\b_1$ takes the values 0.58946, 0.51117, 0.50015, 0.50000 at $t=$
1, 2, 3, 4 steps
 respectively. \par
The asymptotic value of $\b_1$ at $t\rightarrow \infty$ is 1/2. It can be
explained as follows.
 For bare action with $\th^b=\pi $ and $\{\b^b_q\}=0 (q=$ all),
 bare F-function  is given by $F^b(x)=\exp(ix/2)$.
It can be expanded by IR characters as $\Sigma\c_q \Ft_q$ with
$\Ft_q=(-1)^q/(\a-q)\pi$ (where $\a=1/2$ for $\th^b=\pi$ ).
After $t$-steps of RG transformations, the partition function becomes
$$F^{(t)}(x)=\sum\c_q(x) \Ft_q^{\l^{2t}}, (\l>1)\eqno(27).$$
For $\a=1/2$, there is a symmetry
$$\Ft_{1-q}=\Ft_q, \hskip 1cm (q= {\rm any \ \ integer)}.   \eqno(28)$$
Since $|({1\over 2}-q)\pi|>1$ for any integer $q$,
$$1\gg \Ft^{\l^{2t}}_1=\Ft^{\l^{2t}}_0\gg\Ft^{\l^{2t}}_2=\Ft^{\l^{2t}}_{-1}
        \gg\Ft^{\l^{2t}}_3=\Ft^{\l^{2t}}_{-2}\gg \cdots \eqno(29)$$
as $t \rightarrow $ large.  \par
At large $t$,
$$F^{(t)}(x)\cong \c_0(x)\Ft^{\l^{2t}}_0+\c_1(x)\Ft^{\l^{2t}}_1\propto
(\c_0(x)+\c_1(x)), \eqno(30)$$
due to eq.(28). Then real action is given by fixed point form $\b_q=
(-1)^{q-1}/2|q|, (q\not=0),$
i.e., $\b_1=1/2, \b_2=-1/4, \b_3=1/6, \cdots$ and the imaginary action is
given by $\g_q=i(-1)^q/2q, (q\not=),$ i.e., $\g_1=-i/2, \g_2=i/4, \g_3=-i/6,
                                         \cdots$ .
These fixed point coupling constants come from
$$ \eqalign{ F^{(t)}(x) &\propto \c_0(x)+\c_1(x)=1+e^{i \ x} \cr
        &=2\cos {x\over 2}\exp (i{x\over 2})
              =\exp\{\ln(2\cos{x\over 2})+{ix \over 2}\}\cr
        & =\exp \{\sum_q \c_q(x){(-1)^{q-1}\over 2|q|}
               +i\sum_q \c_q(x){(-1)^q\over 2q} \}.\cr}\eqno(31) $$
This finite value of the infrared fixed point in $\th^b=\pi$ case
is exceptional. For any other $\th^b(\not= \pi)$,  coupling
constants  will converge to $\{\b_q\}, \{\g_q\} \rightarrow 0$
as $t\rightarrow \infty$. It can also be seen analytically in the case $\th
 \ltsim \pi$,
i.e., $\a={1\over 2}-\ve( \ve >0)$ as follows.
It gives $\Ft_0=1/({1\over 2}-\ve)\pi\gtsim \Ft_1=1/({1\over 2}+\ve) \pi$
and
$$ F^{(t)}(x)\sim \c_0(x)\Ft^{\l^{2t}}_0
    +\c_1(x)\Ft^{\l^{2t}}_1 \rightarrow
     \c_0(x)\Ft^{\l^{2t}}_0={\rm const(real),} \eqno(32)$$
as $t \rightarrow \infty, $ namely, $\b_q=0$ and $\g_q=0$
for all $q$ at infrared fixed point.\par
\vskip 1cm
%
\newsec{Conclusion} 
Real space renormalization group analysis is made about U(1) gauge theory with
 $\th$-term. We  found phase transition at $\th = \pi $ leading to
deconfinement phase because at this point coupling constant is not controlled
 by strong coupling infrared fixed point. The appearance of the phase
 transition is consistent with Monte Carlo analysis( Wiese) and analytical
argument( Coleman). Namely $\th $-term leads to constant background
electric field and the strength given by  $\th $-term is large enough to
make charged particle become deconfined. \par
It will be quite interesting to perform similar analysis (1)in 3 dimensional
U(1) lattice gauge theory with $\th $-term( Chern-Simons term), and
(2)in two dimensional CP$^{N-1}$ theory with $\th$-term.\par
\vskip 1cm
%
%
\bigbreak\bigskip
\centerline{{\bf Acknowledgments}}\nobreak
The authors thank   Harada, Tominaga, Tsuzuki and particle physics group
in Kyushu University for valuable discussions.\par
Hassan is grateful to Ministry of Culture and Education of Japan for the
Scholarship.\par
%
%
%
\bigskip
%
\footatend\vfill\supereject\immediate\closeout\rfile\writestoppt
\baselineskip=14pt\centerline{{\bf References}}\bigskip{\frenchspacing%
\parindent=20pt\escapechar=` \input refs.tmp\vfill\eject}\nonfrenchspacing
\leftline\bf{FIGURE CAPTIONS}
\item{Fig. 1.}RG flow diagram of coupling constants of real action projected
              onto $(\b_1, \b_2)$ plane. Gaussian action is shown by a
              broken line with slope $ -1/4$.
              $\th^b=0.9\pi $. Various $\b^b$'s lead to single
              trajectory.\par
\item{Fig. 2.}RG flow diagram of coupling constants of imaginary ($\th$-term)
              action projected onto $({\rm Im} \g_1, {\rm Im} \g_2)$ plane.
                $\th^b=0.9\pi$. Various $\b^b$'s lead to single trajectory.\par
\item{Fig. 3.}RG flow diagram in strong coupling regions projected onto
               $(\b_1, \b_2)$ plane. $\th^b=0.9\pi$. All the trajectories
               starting from various $\b^b$'s lie on a single trajectory.\par
\item{Fig. 4.}RG flow in $({\rm Im} \g_1, {\rm Im} \g_2)$  plane.
               $\th^b=0.9\pi$. \par
\item{Fig. 5.}$\th^b=\pi$ is the exceptional case. Infrared fixed point is not
              $ \b^b=0 \ (q=$ all), but $\b_1=1/2, \ \b_2=-1/4$ etc. Even
              starting from $\{ \b^b_q \}=0 \  (q=$ all), we obtain $\b_1\sim
              1/2, \ \b_2\sim-1/4$ etc. after several steps of
               RG transformations.\par
\item{Fig. 6(a).}RG flow for different $\th^b$ near $\pi$. Each trajectory is
                specified by $\th^b$'s. The trajectories
               with the same $\th^b$ converge to single trajectory
               irrespective of $\b^b$'s. For $\th^b$ very close to $\pi$,
               e.g. $\th^b=0.999\pi$, the trajectories starting from $\b^b=$
               large(weak coupling) rapidly go to the point near to the fixed
               point $ (\b_1=1/2, \b_2=-1/4, \ \cdots)$ of $\th^b=\pi$
               and stay for long time around there and finally start to
               infrared fixed point $\b_q=0\ (q=$ all) of $\th^b\not=\pi$.
                   \par
\item{Fig. 6(b).}Flow in $(\b_1, {\rm Im}\g_1)$ plane. All $\th^b(\not=\pi)$'s
               are governed by an infrared fixed point $\{\b_q\}=0\ (q=$ all)
              but $\th^b=\pi$ is governed by the fixed point $\b_q\not=0$.
              $\b^b=4.0$ for all cases.  \par
\item{Fig. 7.}Gell-Mann-Low function($t_G$) plotted against $\th^b$. It
              clearly shows the phase transition at $\th^b=\pi$.\par
\item{Fig. 8.}Convolution of two plaquettes.\par
\bye